\documentstyle[sprocl,12pt]{article}
\input{psfig.sty}
\begin{document}
\title{Ultrafast Coherent Spectroscopy of the Fermi Edge Singularity}
\author{\underline{Diego Porras}, J. Fern\'andez-Rossier and C. Tejedor}
\address{Departamento de F\'{\i}sica Te\'orica de la Materia Condensada.
Universidad Aut\'onoma de Madrid. Cantoblanco, 28049 Madrid. Spain}  

\maketitle

\begin{abstract}

In this work we present a theoretical description of the transient response of
the Fermi Edge Singularity (FES). We study the linear and the nonlinear
response of an n-doped QW to laser pulses in the Coherent Control (CC) and Four
Wave Mixing (FWM) Configurations. By means of a bosonization formalism we
calculate the FWM signal emitted by the sample when it is excited by pulses
spectrally peaked around the FES  and we show that the long time behavior of
the nonlinear signal is very similar to the linear case.

\end{abstract}


\section{Motivation}

The promotion of an electron from a localized state in a valence band (VB) to
an empty state in a partially filled conduction band (CB) is accompanied by a
dynamical response of the Fermi gas. The enhancement of the  absorption
probability when the new electron is promoted just above the Fermi energy is
known as the Fermi Edge Singularity (FES)\cite{Mahan}. This phenomena has been
experimentally observed, by means of continuous wave  spectroscopy, both in
metals\cite{Mahan} and doped semiconductors \cite{doped}.  FES is understood as a 
consequence of the appearance of a {\em coherent} gas of low energy collective
excitations of the Fermi sea  due to both the sudden appearance of the
localized  hole and the new electron. It has been shown \cite{Schotte} that
these collective excitations can be considered as Tomonaga bosons which are in
a Glauber coherent state.

In this paper we propose two types of transient coherent experiments, Coherent
Control (CC) and Four Wave Mixing (FWM) in doped Quantum Wells, in order to 
address the following questions: {\em i)} Can be these Tomonaga bosons
coherently controlled? {\em ii)}  How is the decay of the Tomonaga bosons
coherence?  and {\em iii)} What  is the nonlinear response of the  FES ? In
this context, pump and probe  experiments have been suggested for the transient
spectroscopy of the FES \cite{Chemla}. The experience acquired  in the case of 
undoped samples shows that both  CC \cite{CC} and FWM  \cite{FWM} are   precise
and conceptually simple tools to probe  excitonic coherence and nonlinearity.
We propose to  excite an n-doped QW with laser pulses  which promote
electrons from localized states in the VB to the Fermi level in the CB. The
spectral width  of the laser pulses ($1/\Delta t$ ) must be much smaller than
the Fermi energy (measured with respect the bottom of the
CB), $\epsilon_F$,  so that the only intraband excitations are the
Tomonaga bosons. Therefore, our theoretical approach is not valid to address
the  FWM experiment of Kim {\em et al.} \cite{Kim} but it could be compared to the
zero magnetic field case of the experiments of Bar-Ad {\em et al.} \cite{Bar}.

\section{Hamiltonian and Linear Response: Coherent Control}

The problem of the linear response function of the FES was solved long time ago
in a  particularly simple way by a description of the Fermi Sea excitations in
terms of  bosonic operators\cite{Schotte}. If we consider that the photoexcited
hole creates a  spherical localized potential then the Hamiltonian is:
\begin{eqnarray}
H = E_0 \ d^ \dagger d + \sum_{k=0}^{k_D} \epsilon_k a^{\dagger}_k
a_{k} + \frac{V}{N} \sum_{k,k^{\prime}=0}^{k_D}
a_k^{\dagger}a_{k^{\prime}}d^{\dagger}d     
\end{eqnarray}
where ${E_0}$, ${d^{\dagger}}$ are the energy and creation operator of the
localized hole, ${a_k^{\dagger}}$ are the creation operators of the electrons
in the conduction band and ${k_D}$ is the bandwidth \cite{Schotte}. Here we
will only consider s-wave scattering, and reduce the initial problem to a 1-D
one, so that the index k corresponds to the radial wave function index.  The
polarization induced in the sample by an electric field is described by the
operator ${P(t)= \mu \ a^{\dagger} d^{\dagger} + h.c.}$, where
${a=\sum_{k=0}^{k_D}a_k}$ and ${\mu}$ is the dipole matrix element.  The mean
value of P(t) in the absence of inelastic scattering (decoherence)  can be
calculated by means of the linear response function. For zero temperature
($T=0$) we have:
\begin{eqnarray}
\label{lineal}
\chi^{(1})(t) \propto -i \ \mu^2 \theta(t)
(i \epsilon_F t)^{\alpha} e^{-i\epsilon t}+h.c.
\end{eqnarray}
where  ${\rho}$ is the density of states at the Fermi surface, ${\epsilon}=E_g
+ \epsilon_F  -(V\rho)^2\epsilon_F $ is the FES resonance energy, $E_g$ is 
the semiconductor gap and ${\alpha}=(1+V\rho)^2$ is the exponent of the
singularity. Let us consider now the  CC of the linear response of the sample.
The signal emitted is proportional to the polarization created by the incident
laser. We will work in the Rotating Wave Approximation (RWA). If the laser
pulse has the form ${E(t) = E_A(t) + E_B(t) + h.c.}$, with
${E_A(t)={\mathcal{E}}_A (t) e^{-i \epsilon t}}$, ${E_B(t) = E_A(t-\tau)}$, and
we approximate  ${{\mathcal{E}}_A (t)}$ by a delta function, then this signal
is (for ${t>>\tau}$):
\begin{eqnarray}
\label{linealcc}
F^{(1)} \propto - i (i \epsilon_F t)^{\alpha} (1+e^{i \epsilon \tau}) + h.c
\end{eqnarray} 
This quantity can be measured by means of the time integrated luminescence, as
in the undoped case \cite{CC}, and it turns out to be  $I \propto |F^{(1)}|^2
\propto 1+Cos(\epsilon \tau)$. Therefore, CC of the FES is possible.  This can
be understood as the interference of the electric dipole induced by the first
pulse with that induced by the second pulse. Microscopically, Tomonaga bosons
are being coherently controlled in much the same way than the CC of optical
phonons \cite{Wehner}.

\section{Four Wave Mixing signal}

In FWM experiments, the system is excited by two phase locked pulses having 
momenta ${\bf k_A}$ and ${\bf k_B}$ with $|{\bf k_A}|=|{\bf k_B}|$ and a 
relative delay $\tau $. The excitation of electrons near the Fermi surface can
be achieved by choosing pulses spectrally peaked around the FES resonance.   
The FWM signal comes from the interference between  the polarization created by
the first pulse ($t=0$) and that of the second pulse  ($t=\tau$). We  neglect
the (exponential) decay of the polarization  due to inelastic scattering and to 
inhomogeneous broadening of the holes. We  show that there is  an  {\em
intrinsic  potential} decay, due to the dephasing of the Tomonaga Bosons, which
could be observed in favorable conditions.

To calculate the intrinsic decay we  need to obtain the third-order contribution
to the polarization induced in the sample, which is proportional to the FWM
signal emitted in the $2{\mathbf{k}_B}-{\mathbf{k}_A}$ direction. We consider
two pulses ${E_{A,B}(t)={\mathcal{E}}_{A,B} (t)e^{-i\epsilon t}}$, where
${{\mathcal{E}}_{A,B}(t)}$ are the amplitudes of the pulses in the directions
${\bf k_A}$, ${\bf k_B}$ . The FWM signal can be written in terms of the
third-order response function:
\begin{eqnarray}
& F(t)=&  \int _{- \infty} ^t dt_1 dt_2 dt_3 \     
\chi^{(3)} (t-t_1,t-t_2,t-t_3) \  
E^{\ast} _A(t_2)E_B(t_1)E_B(t_3)
\end{eqnarray}
By extending up to third order the usual perturbation scheme
developed for obtaining the Kubo formula, one obtains:  
\begin{eqnarray}
\chi^{(3)} (t)  & = & -i  \ \ [  
\ \theta (t-t_1)\theta (t_1-t_2)\theta (t_2-t_3) \ \langle P(t) P(t_1)P(t_2)P(t_3)
\rangle 
\nonumber \\ & &  \qquad \qquad + \ \ 
\theta (t-t_1)\theta (t_1-t_2)\theta (t_2-t_3) \ \langle P(t_2)P(t_1)P(t)P(t_3) \rangle 
\ \ ]
 +h.c. 
\label{nlsignal}
\end{eqnarray}
where the brackets $\langle \rangle$ mean thermal averages. Equation
(\ref{nlsignal}) has been obtained within the RWA  in which the width $\Delta
t$ of each pulse is much larger than the  period ${\propto \epsilon^{-1}}$ of
the exciting light. Here we will also consider, for simplicity, delta pulses:
${E_A(t) = E_A \delta (t) e^{-i\epsilon t}}$, ${E_B(t) = E_B \delta (t-\tau )
e^{-i\epsilon (t-\tau )}}$. This assumption does not imply a contradiction with
the RWA: it is justified if the variations of the correlation functions are
slow enough in a time interval ${\Delta t}$.  In (\ref{nlsignal}) only the
second term takes a non-vanishing value. Our main task now is the calculation
of the correlation function, for which the bosonization formalism is extremely
useful. If we define ${H_i}$, ${H_f}$ as the Hamiltonian without and with hole,
respectively, then:
\begin{eqnarray}
\langle P(t_2)P(t_1)P(t)P(t_3) \rangle = \mu^4
\langle e^{i H_i t_2}ae^{-i H_f (t_2-t_1)}a^{\dagger}e^{-i H_i (t_1-t)}ae^{-i H_f (t-t_3)}
a^{\dagger}e^{-i H_i t_3} \rangle 
\label{propagator}
\end{eqnarray}
If we assume that $\Delta t>>(\epsilon_F)^{-1}$, only excitations near the
Fermi Edge will be relevant and we can use  the  Schotte \cite{Schotte}
approach to  reexpress this correlation function in the bosonization formalism:
 \begin{eqnarray}
\langle P(t_2) P(t_1)P(t)P(t_3) \rangle \propto
\langle B^{\dagger} (t) B(t_1) B^{\dagger} (t_2) B(t_3) \rangle e^{-i\epsilon t} 
e^{i \epsilon (t_1-t_2+t_3)} 
\end{eqnarray}
where ${B^{\dagger}}$ are operators that create a coherent state of Tomonaga
bosons $b^{\dagger}_k$:
\begin{eqnarray}
B(t) = exp \left[ \sum_{k=0}^{k_F} \frac{(1+V\rho)}{\sqrt{kN}} 
\left( b^{\dagger}_k e^{i \frac{k}{\rho} t} - b_k e^{-i \frac{k}{\rho} t}
\right) \right]    
\label{coherent}
\end{eqnarray}
 After  integration with the pulses we arrive at:
\begin{eqnarray}
F^{(3)}(t) \propto -i \mu^4 e^{-i \epsilon (t-2\tau )} \theta(t-\tau) \theta(\tau)
\langle B^{\dagger} (0) B(\tau) B^{\dagger} (t) B(\tau) \rangle 
\end{eqnarray}
We consider a bath of bosons in equilibrium at temperature T (with $k_B T <
\epsilon_F$) for the thermal average:
\begin{eqnarray}
F^{(3)}(t)  \propto  -i \ \mu^4 \ \theta(t-\tau) \theta(\tau) \ e^{-i \epsilon (t-2\tau )} \ 
exp \left( \alpha \sum_{k=0}^{k_F} (\frac{1}{2}+N_B) \frac{1}{kN} | 1-2e^{i \frac{k}{\rho} \tau}
+ e^{i \frac{k}{\rho} t} |^2 \right) 
\nonumber \\ \times exp \left( i \alpha \sum_{k=0}^{k_F} \frac{1}{kN} 
\left( -2 sin \frac{k}{\rho} \tau + sin \frac{k}{\rho} t \right) \right)    
\label{fwmsignal}
\end{eqnarray}

where $N_B=(e^{\frac{k/{\rho}}{ k_B T}}-1)^{-1} $. We can express the sum in
momentum space as an integral: ${\sum^{k_F}_{k=0}}$ = ${\int^{k_F}_0 dk}$.
If we consider the limit ${t>>\tau}$ we obtain a very simple expression for the
complete FWM signal:
\begin{eqnarray}
F^{(3)}(t,T)  \propto  -i \ \mu^4 \ \theta(t-\tau) \theta(\tau) 
\ e^{-i \epsilon (t-2\tau )} \ 
\left[ \left( \frac{i\epsilon_F}{\pi k_B T} \right)^3 
sinh^2(\pi k_B T \tau) \ sinh(\pi k_B T t) \right]^{\alpha} 
\label{FWMT}
\end{eqnarray}
This signal has the zero temperature limit:
\begin{eqnarray}
F^{(3)}(t,T=0)  \propto  -i \ \mu^4 \ \theta(t-\tau) \theta(\tau) 
\ e^{-i \epsilon (t-2\tau )} \ 
(i\epsilon_F \tau)^{2\alpha}  (i\epsilon_F t)^{\alpha}
\label{FWMT0}
\end{eqnarray}

\section{Conclusions}

In this paper we have shown that temporal CC of the Tomonaga bosons can be
achieved.  We have also shown that the bosonization formalism is a simple tool
to calculate  the nonlinear susceptibility $\chi^{(3)}$, for both zero and
nonzero temperature. We have calculated the FWM signal, for the case of Delta
pulses, and we have shown that $F^{(1)}$ and $F^{(3)}$ have an {\em identical} 
dependence on the detection time $t$ for both $T=0$ and $T\neq 0$
\cite{Ohtaka}. Neglecting inelastic scattering  $F^{(3)}$ has  an intrinsic 
{\em potential} decay as a function of $\tau$. It must be pointed out that this
potential decay will be superimposed, in real experiments, to an exponential
decay produced by   electron-electron, electron-phonon interactions and 
inhomogeneous broadening of the holes.

\section{Acknowledgements}

Work supported in part by MEC of Spain under contract PB96-0085, Fundacion
Ram\'on Areces and CAM under contract 07N/0026/1998.  Diego Porras thanks
Spanish Education Ministry for its FPU grant.


\end{document}